\documentclass[letterpaper, fontsize = 10 pt]{IEEEtran} 
\pdfoutput=1
\usepackage[T1]{fontenc} 
\usepackage{fourier} 
\usepackage[english]{babel} 
\usepackage{amsmath,amsfonts,amsthm,amssymb} 

\usepackage{amsmath}
\usepackage{mathtools}
\usepackage{amssymb}
\usepackage{graphicx}
\usepackage{epstopdf}
\usepackage{color}
\usepackage{secdot}
\usepackage{cite}
\usepackage{tikz}
\usepackage{pgfplots}
\usepackage{authblk}
\usepackage{subcaption}
\usepackage{enumerate}
\usepackage{multirow}
\usepackage{float}
\usepackage{dblfloatfix}
\usepackage{soul}

\newtheorem{theorem}{Theorem}

\newcommand{\red}[1]{{\color{black}#1}}

\setstcolor{red}

\begin{document}
	
	\title{A Generalized Framework for Simultaneous Long-Short Feedback Trading}
	%
	%
	%
	
	\author{Joseph~D.~O'Brien,
		Mark~E.~Burke,
		and~Kevin~Burke
		\thanks{\hspace{-.25cm}Joseph D. O’Brien, Kevin Burke, and Mark E. Burke are with MACSI, Department of Mathematics and Statistics, University of Limerick, V94 T9PX Limerick, Ireland (e-mail: joseph.obrien@ul.ie; kevin.burke@ul.ie; mark.burke@ul.ie).}.
		\thanks{\hspace{-.25cm}This work was funded by Science Foundation Ireland Grant 16/IA/4470 (J.D.O'B.).}
	}
	
	%
	%

	\markboth{}%
	{Shell \MakeLowercase{\textit{et al.}}: Bare Demo of IEEEtran.cls for IEEE Journals}
	%



	\maketitle
	\begin{abstract}
		We present a generalization of the Simultaneous
		Long-Short (SLS) trading strategy described in recent control
		literature wherein we allow for different parameters across the
		short and long sides of the controller; we refer to this new
		strategy as Generalized SLS (GSLS). Furthermore, we investigate
		the conditions under which positive gain can be assured within
		the GSLS setup for both deterministic stock price evolution and
		geometric Brownian motion. In contrast to existing literature
		in this area (which places little emphasis on the practical
		application of SLS strategies), we suggest optimization procedures for selecting the control parameters based on historical
		data, and we extensively test these procedures across a large
		number of real stock price trajectories (495 in total). We
		find that the implementation of such optimization procedures
		greatly improves the performance compared with fixing control
		parameters, and, indeed, the GSLS strategy outperforms the
		simpler SLS strategy in general.
	\end{abstract}
	

	\section{Introduction\label{sec:intro}}
	
	The use of feedback in a control-theoretic scenario has been well studied within a variety of different fields -- its use dates back at least two millennia where the flow of water was regulated to improve the accuracy of water clocks. The application of feedback models became widespread during the Industrial Revolution and since then their use has become ubiquitous in engineered systems \cite{kumar2014control}. In recent years, the methodology has been applied in the setting of equity trading wherein a closed-loop feedback system is used to modulate the investment level, $I(t)$, in response to changes in the equity price, $p(t)$. As this basic system reacts to changes in price, rather than making predictions about future price movements, the resulting strategy is often described as ``model free''.
	
	Our starting point is the trading framework developed by Barmish which initially considered only a long investment~\cite{barmish2008trading}, but was later extended so that one takes both long and short positions in the equity -- the so-called ``simultaneous long-short'' (SLS) strategy \cite{barmish2011performance}. (For related earlier work, see~\cite{cover1991universal,dokuchaev1998positive,dokuchaev1998arbitrage,dokuchaev2002bounded,dokuchaev2004universal}.)
	The key feature of the SLS strategy is that the resulting gain function, $g(t)$, is guaranteed to be positive under (potentially restrictive) assumptions of deterministic $p(t)$ (albeit this is not assumed to be known), continuous trading, perfect liquidity, and the absence of transaction fees (for more details see \cite{barmish2011performance}). Further developments in this area included the consideration of interest rates and collateral requirements \cite{barmish2011arbitrage, barmish2012market, barmish2015stock, malekpour2013stock}. This work culminated in~\cite{barmish2016new} which laid the foundations for many future research directions such as using a controller with delay \cite{malekpour2016stock}, different price process models \cite{baumann2017stock, baumann2017simultaneously}, time varying price evolution parameters \cite{primbs2017robustness}, \red{assets with correlated prices \cite{deshpande2018generalization}}, and also the use of a proportional-integral (PI) controller rather than the proportional controller which was used in the original SLS strategy \cite{malekpour2018generalization}.
	
	While various extensions to the SLS framework have been proposed to date, the basic underlying model structure has remained essentially unchanged in the sense that the initial investment and feedback parameters are the same for both the long and short investments. However, a real trader may wish to tune these components of the controller in different ways. We also note that in the current literature, although much theoretical work has been done (mainly on investigating the positive gain property under different scenarios), testing such SLS strategies in practical situations has been much more limited. Specifically, testing has been typically carried out on a very small number of stock price series (one or two), and, hence, there is very little sense of the general performance of these control-based strategies. (Interestingly however, various other technical trading strategies were examined in \cite{hsu2005reexamining}, albeit using only four market index series.) Furthermore, there has also been a lack of guidance as to how one should select the feedback parameter in practice, where, apparently, this choice has been quite arbitrary in applications shown to date. \red{This is despite \cite{barmish2016new} highlighting that ``the most important new class of research problems $\ldots$ involves optimal gain selection'', recognizing that the feedback parameter has a ``large effect [on] the mean and variance of the gain'' (see also \cite{baumann2017stock}).} With the level of gain being quite modest in some applications (as was mentioned in \cite{primbs2017robustness,malekpour2018generalization}), one wonders whether or not greater \red{and/or more stable} gains can be made through a more objective selection process, e.g., by optimizing some criterion.
	
	The aim of the current article is to tackle the issues raised in the previous paragraph by:
	\begin{enumerate}[a)]
		\item generalizing SLS to allow parameters which differ across the short and long side -- the resulting strategy we call ``Generalized SLS'' (GSLS) -- thereby permitting greater flexibility beyond the classical SLS approach;
		\item proposing an optimization procedure for selecting control parameters based on historical data, providing an objective process for their selection;
		\item extensively analyzing the performance of GSLS (and SLS) with our proposed optimizer on a large number (495) of real stock price trajectories to determine the practical usefulness of the control-based trading concept.
	\end{enumerate}
	The rest of this article is set out as follows. In Section \ref{sec:SLS} we introduce the classical SLS strategy of \cite{barmish2016new}, after which, in Section \ref{GSLS}, we extend to our proposed GSLS strategy, deriving a number of new analytical results. Section \ref{Optim} provides a suggestion on how one might objectively choose control parameters within GSLS (and, hence, SLS and other varieties thereof) which is then tested on historical data in Section \ref{sec:testing}. Finally, we conclude with some discussion in Section \ref{sec:discussion}.
	
	\section{Simultaneous Long-Short (SLS) Strategy\label{sec:SLS}}
	
	As is customary, we assume that the strategy is applied in the setting of an ``idealized market'' where the primary assumptions are: (i) continuous-time trading is possible, (ii) the equity is perfectly liquid so that shares can be purchased with no gap between the bid and ask prices, and (iii) there are no transaction fees or interest rates. See~\cite{barmish2011performance} for further details.
	
	We now consider the key components of the SLS strategy. Let $p(t)$ represent the price of the equity at time $t \geq 0$. Furthermore, let $I(t)$ be the level of investment at time $t$, and $g(t)$ be the gain function, i.e., the cumulative trading gain over the period $[0,t]$ where $g(t) < 0$ represents an overall trading loss; also note that, by definition, $g(0) = 0$. In the simpler long strategy of \cite{barmish2008trading} (i.e., no short component), the amount invested is given by
	
	\begin{equation*}
	I(t) = I_0 + Kg(t),
	\end{equation*}
	where $K \ge 0$ is the feedback parameter and $I_0>0$ is the initial investment. The gain made on a given stock is the amount invested in this stock multiplied by the percentage change in stock price, giving the incremental gain equation
	
	\begin{equation*}
	dg = \frac{dp}{p}I = \frac{dp}{p}(I_0 + Kg)
	\end{equation*}
	whose solution yields
	
	\begin{equation*}
	g(t) = \frac{I_0}{K}\left\{q(t)^K - 1\right\},
	\end{equation*}
	where $q(t) = p(t)/p(0) > 0$ is the ratio of the current and initial stock price, and, hence, we see that the investment level is $I(t) = I_0 q(t)^K$.
	
	The SLS strategy extends the above (but follows along the same lines) by introducing simultaneous long and short investments, $I_L(t)$ and $I_S(t)$ respectively, and their associated cumulative gain functions, $g_L(t)$ and $g_S(t)$. The strategy is defined by
	
	\begin{align*}
	I_L(t) &= I_0 + Kg_L(t), \\
	I_S(t) &= -I_0 - Kg_S(t),
	\end{align*}
	where
	
	\begin{align*}
	g_L(t) &= \frac{I_0}{K}\left\{q(t)^K - 1\right\},\\
	g_S(t) &= \frac{I_0}{K}\left\{q(t)^{-K} - 1\right\}.
	\end{align*}
	
	Note that the overall investment in this strategy is given by $I(t) =~I_L(t) + I_S(t) =~K\{g_L(t) - g_S(t)\}$ and, in particular, that $I(0) = 0$,  i.e., the simultaneous long and short positions (which are equal and opposite when $t=0$ with magnitude $I_0$) lead to an initial net zero investment position. As shown in \cite{barmish2011performance}, under the idealized market, the cumulative gain at time $t$ for this SLS strategy is
	
	\begin{equation}
	\label{totalgain}
	g(t) = g_L(t) + g_S(t) = \frac{I_0}{K}\left\{q(t)^{K} + q(t)^{-K} -2\right\},
	\end{equation}
	and, hence, $g(t) > 0$ provided that  $q(t) \ne 1$, i.e., this strategy always makes a profit except when $p(t) = p(0)$.

	\section{Generalized SLS Strategy\label{GSLS}}
	
	As seen in Section \ref{sec:SLS}, the basic SLS strategy is composed of the simpler long strategy with the addition of a short component. In particular, the short component mirrors the long component in the sense that it shares the same initial investment, $I_0$, and feedback parameter, $K$. However, we now view this as an interesting special case of a more general framework in which the long and short components have distinct parameters as follows:
	\begin{align*}
	I_L(t) &= I_0 + K g_L(t), \\
	I_S(t) &= -\alpha I_0 -\beta K g_S(t),
	\end{align*}
	where $\alpha,\beta>0$, and
	\begin{align*}
	g_L(t) &= \frac{I_0}{K}\left\{q(t)^K - 1\right\},\\
	g_S(t) &= \frac{\alpha I_0}{\beta K}\left\{q(t)^{-\beta K} - 1\right\}.
	\end{align*}
	We refer to this as the Generalized SLS (GSLS) strategy, within which $\alpha = \beta = 1$ corresponds to SLS, and whose overall gain function is
	\begin{equation}
	g(t) = \frac{I_0}{K} \left[q(t)^K - 1 + \frac{\alpha}{\beta}\left\{q(t)^{-\beta K} - 1\right\}\right]
	\label{ComboEq}
	\end{equation}
	Notwithstanding the guaranteed positive gain property of the important SLS special case, in practice, this specific strategy cannot uniformly yield the optimal gain. We will discuss this in the following sections.
	
	\subsection{GSLS with $\beta=1$\label{sec:gslsb1}}
	
	Before investigating the GSLS strategy, it is instructive to first consider the special case where $\beta=1$, i.e., the case where the long and short initial investments differ, but the feedback parameter is the same in both components. In this specific case note that the gain is given by
	\begin{equation}
	g(t) = \frac{I_0}{K} \left[q(t)^K - 1 + \alpha\left\{q(t)^{-K} - 1\right\}\right],
	\label{beta1}
	\end{equation}
	where we see that $g(t) = 0$ when $q(t) = 1$ or when 
	$q(t) = \alpha^{1/K}$. Furthermore, viewed as a function of $q(t)$, the gain function has a single turning point between the two roots at $q(t) = \alpha^{1/(2K)}$; it is a global minimum of
	$-I_0 (\alpha^{1/2} - 1)^2/K < 0$. Note that in the classic SLS case, where $\alpha=1$, these three points coincide to become a single root and global minimum of zero at $q(t) = 1$, yielding the associated positive gain property.
	
	Assume that $\alpha < 1$, so that we invest in the short component to a lesser extent than the long component. In this case, $g(t) > 0$ provided that $q(t) \not\in [\alpha^{1/K}, 1]$. This can be seen as a trade-off between risk and reward. First consider the classic SLS strategy where $\alpha = 1$. This is, theoretically, risk free in the sense that $g(t) > 0$ provided that $q(t) \ne 1$. On the other hand, in selecting $\alpha < 1$ and, hence, investing more on the long side, we are essentially admitting a belief that the stock price will increase in the future, i.e., $q(t) > 1$. Indeed, it is easy to show that $g_{\alpha<1}(t) > g_{\alpha=1}(t)$ when the price rises ($q(t) > 1$) so that the reward goes up by choosing $\alpha < 1$ -- but, of course, the risk also goes up since there is a chance that $g(t) < 0$.
	
\begin{figure}[h]
		\centering
		\begin{tikzpicture}[scale = 1]
		\begin{axis}[
		axis lines = center,
		xlabel=$q$,ylabel=$g$,
		%
		y label style={at={(axis description cs:-0.1,1)},anchor=north},
		x label style={at={(axis description cs:1,0.05)},anchor=east},
		legend style={at={(0.1,1.0)},
			anchor=north west,legend columns=1},
		domain=0.5:2,
		legend style = {nodes={scale=0.5, transform shape}}]
		\addplot[samples = 100, blue, smooth, thick] {(1/1)*(x^1 - 1 + (0.5/1)*((x)^(-1*1) - 1))};
		\addplot[samples = 100, blue, dashed, thick] {(1/3)*(x^3 - 1 + (0.5/1)*((x)^(-1*3) - 1))};
		\addplot[samples = 100, red, smooth, thick] {(1/1)*(x^1 - 1 + (1/1)*((x)^(-1*1) - 1))};
		\addplot[samples = 100, red, dashed, thick] {(1/3)*(x^3 - 1 + (1/1)*((x)^(-1*3) - 1))};
		\addplot[samples = 100, green, smooth, thick] {x-1};
		\legend{$\alpha=0.5,\ \beta=1,\ K= 1$\\$\alpha=0.5,\ \beta=1\, K= 3$\\$\alpha=\beta=1,\ K= 1$\\$\alpha=\beta=1,\ K= 3$\\ ``going long''\\};
		\end{axis}
		\end{tikzpicture}
		\caption{Plots of the gain for various values of $q(t)$ when the trading strategy is GSLS ($\alpha=0.5$, $\beta=1$) (blue), SLS (red), and simply going long (green). Furthermore, for GSLS and SLS both $K=1$ (solid) and $K=3$ (dashed) cases are shown.\label{fig:gslsvlong}}
	\end{figure}
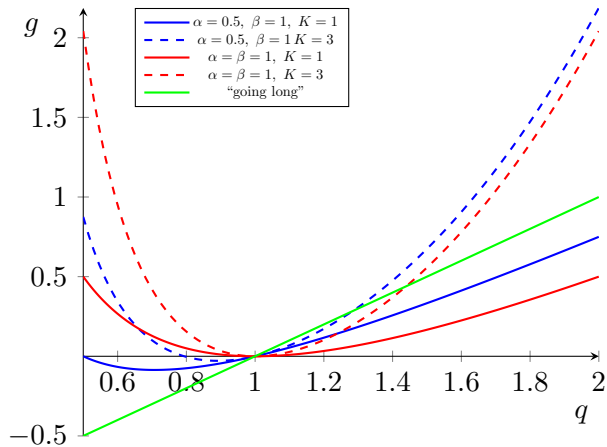

	Now consider a more standard trading strategy (i.e., not control-based) where we simply go long, i.e., buy at time zero and sell at time $t$. In this case, the investment is fixed at $I(t) = I_0$ and $g(t) = I_0\left\{q(t) - 1\right\}$. In a similar manner to GSLS with $\alpha < 1$, we are anticipating a price rise. However, we are guaranteed to make a loss if $q(t) < 1$, whereas, with GSLS, we can still profit if $q(t) < \alpha^{1/K}$. In other words, we are reducing the risk compared with simply going long. Furthermore, although with GSLS the risk is reduced relative to going long, the gain is not uniformly lower (for all control parameter values) even when $q(t) > 1$ as demonstrated by Figure \ref{fig:gslsvlong}.

	The above discussion relates to the case where $\alpha < 1$. When $\alpha > 1$, we have that $g(t) > 0$ provided that $q(t) \not\in [1, \alpha^{1/K}]$, and, moreover, we can show that $g_{\alpha>1}(t) > g_{\alpha=1}(t)$ when the price falls ($q(t) < 1$). Therefore, in general, the choice of $\alpha=1$ is sub-optimal in the sense that, if one possesses knowledge on the likely direction of the stock price, a greater return can be made by choosing $\alpha \ne 1$. In practical situations, a trader is likely to require some input into his/her trading strategy such that GSLS may be more attractive than SLS. Furthermore, GSLS may be seen as lying somewhere between SLS and standard trading in that trader knowledge can be incorporated as in standard trading, but where the risk is reduced as in SLS. Interestingly, since $\alpha^{1/K} \rightarrow 1$ as $K\rightarrow \infty$, the condition for positive gain when $\alpha \ne 1$ is simply that $q(t) \ne 1$ when $K$ is large. This suggests that a large enough $K$ value can overcome a poor $\alpha$ choice; indeed, GSLS with large $K$ essentially behaves as SLS.
	
	\subsection{Positive Gain in GSLS}
	
	In the previous section we discussed GSLS under the restriction that $\beta = 1$. We now consider the more general scenario where $\beta$ is not necessarily equal to $1$. In particular, we investigate the conditions for positive gain within GSLS.
	
	
	\begin{theorem}
		In the idealized market, the gain made by the GSLS strategy is positive $\forall K,\beta$ provided that \mbox{$(1-\alpha)\ln\{q(t)\} \ge 0$}.
	\end{theorem}

	{\bf Proof}:
	First we write the gain function, (\ref{ComboEq}), as
	\begin{equation}
	g(t) = I_0 \frac{q(t)^K - 1}{K} + \alpha I_0  \frac{q(t)^{-\beta K}-1}{\beta K}. \label{eq:gainfn2}
	\end{equation}
	Now, consider the inequality
	\begin{equation}
	q^x \geq 1 + x \ln q \label{eq:ineq}
	\end{equation}
	for $q > 0$ and $x \in \mathbb{R}$, due to $m(x) = q^x-1-x \ln q$ being convex with a global minimum of zero at $x = 0$.
	
	Then, from (\ref{eq:ineq}), setting $x=K$ yields
	\begin{align}
	\frac{q^K - 1}{K} &~\ge~ \ln q \label{eq:ineqgL},
	\end{align}
	while setting $x=-\beta K$ yields
	\begin{align}
	\frac{q^{-\beta K}-1}{\beta K} &~\ge~ - \ln q.\label{eq:ineqgS}
	\end{align}
	Hence, combining (\ref{eq:gainfn2}), (\ref{eq:ineqgL}), and (\ref{eq:ineqgS}), we get
	\begin{equation*}
	g(t) \ge I_0 (1 - \alpha) \ln\{q(t)\}
	\end{equation*}
	and, since $I_0 > 0$, we require that $(1 - \alpha) \ln\{q(t)\} \ge 0$ to ensure that $g(t) \ge 0$. $\square$

	The above result generalizes the positive gain result of SLS which we recover by setting $\alpha=\beta=1$. Furthermore, note, in fact, that the result does not depend on the value of $\beta$ (provided that it is finite) so that positive gain is assured once $\alpha=1$ even if the feedback parameters on the short and long side differ. The requirement that $(1-\alpha)\ln\{q(t)\} \ge 0$ can be interpreted as follows: if we expect $q(t) > 1$, we should set $\alpha < 1$, and if we expect $q(t) < 1$, we should set $\alpha > 1$. It is also worth noting that, for large $K$, the behaviour of $g(t)$ is as follows:
	\begin{align*}
	g(t) \sim \left\{
	\begin{array}{lr}
	I_0 q(t)^K \ln\{q(t)\}, & q(t) > 1 \\[0.2cm]
	-\alpha I_0 q(t)^{-\beta K} \ln\{q(t)\}, & q(t) < 1 \\
	\end{array}
	\right.
	\end{align*}
	which is positive in both cases. This suggests that a sufficiently large value of $K$ can yield positive gain irrespective of $\alpha$ and $\beta$. These findings mirror those of Section \ref{sec:gslsb1} which were for GSLS with $\beta=1$. Indeed GSLS with $\beta = 1$ is qualitatively similar to the general case.

	\begin{theorem}
		The GSLS gain function is increasing in the parameters $K$ and $\beta$.
	\end{theorem}
	{\bf Proof}:
	In this case we write the gain function, (\ref{ComboEq}), as
	\begin{equation}
	g(t) = I_0 \frac{q(t)^{K_L} - 1}{K_L} + \alpha I_0  \frac{q(t)^{-K_S}-1}{K_S}. \label{eq:gainfn3}
	\end{equation}
	where $K_L$ and $K_S$ are the feedback parameters on the long and short sides respectively, and the incremental gain equation is given by
	\begin{equation}
	dg = \frac{\partial g}{\partial K_L} dK_L + \frac{\partial g}{\partial K_S} dK_S, \label{eq:ig}
	\end{equation}
	where
	\begin{align*}
	\frac{\partial g}{\partial K_L} &= I_0 \frac{q^{K_L}(K_L \ln q -1)+1}{K_L^2}, \\[0.2cm]
	\frac{\partial g}{\partial K_S} &= \alpha I_0 \frac{q^{-K_S}(-K_S \ln q -1)+1}{K_S^2}. \\
	\end{align*}
	By replacing $x$ with $-x$ in (\ref{eq:ineq}) and multiplying by $q^x$, we can establish another inequality,
	\begin{equation}
	q^x(x \ln q - 1) + 1 \geq 0  \label{eq:ineq2}.
	\end{equation}
	Using (\ref{eq:ineq2}) with $x=K_L$ and $x=-K_S$, respectively, gives $\partial g / \partial K_L \ge 0$ and  $\partial g / \partial K_S \ge 0$. Therefore, the incremental gain, (\ref{eq:ig}), is positive whenever $dK_L$ and $dK_S$ are positive, so that $g(t)$ is increasing in $K_L$ and $K_S$, and, hence, in $K$ and $\beta$ (since $K_L = K$ and $K_S = \beta K$). $\square$
	
	While the above result suggests that a trader can simply increase the feedback parameters to increase profits, this may not be feasible in practice. Firstly, large feedback parameters will cause the controllers (and associated investments) to vary wildly in response to (potentially small) changes in gains which introduces a large degree of variability into the system. Moreover, one of either the long or short investments will become large, and, of course, all traders will have limited resources; for the same reason, one cannot simply increase $I_0$ arbitrarily.
	
	
	
\begin{figure}[h]
		\centering
		\begin{tikzpicture}[scale = 0.9]
		\begin{axis}[
		axis lines = center,
		xlabel=$K$,ylabel=$g$,
		%
		y label style={at={(axis description cs:0.55,1)},anchor=north},
		x label style={at={(axis description cs:1,0.35)},anchor=east},
		legend style={at={(0.65,1.0)},
			anchor=north west,legend columns=1},
		domain=-3:3,
		legend pos  = north west,
		legend style = {nodes={scale=0.5, transform shape}}]
		\addplot[samples = 100, blue, smooth, thick] {(1/x)*((0.5)^x + (0.25/1)*(0.5)^(-1*x) - (0.25/1) - 1)};
		\addplot[samples = 100, yellow, smooth, thick] {(1/x)*((2)^x + (0.25/1)*(2)^(-1*x) - (0.25/1) - 1)};
		\addplot[samples = 100, red, smooth, thick] {(1/x)*((0.5)^x + (2/1)*(0.5)^(-1*x) - (2/1) - 1)};
		\addplot[samples = 100, green, smooth, thick] {(1/x)*((2)^x + (2/1)*(2)^(-1*x) - (2/1) - 1)};
		\legend{$\alpha=0.25,\ q= 0.5$\\$\alpha=0.25,\ q= 2$\\$\alpha=2,\ q= 0.5$\\$\alpha=2,\ q= 2$\\};
		\node[label={},circle,fill,inner sep=2pt, blue] at (axis cs:2,0) {};
		\node[label={},circle,fill,inner sep=2pt, green] at (axis cs:1,0) {};
		\node[label={},circle,fill,inner sep=2pt, red] at (axis cs:-1,0) {};
		\node[label={},circle,fill,inner sep=2pt, yellow] at (axis cs:-2,0) {};
		\end{axis}
		\end{tikzpicture}
		\caption{GSLS gain viewed as a function of $K$ with $\beta =~1$, $I_0 = 1$. The dots are given at $K = \log_{q(t)}(\alpha)$ which is a root of (\ref{beta1}). The cases where this root lies in the negative $K$ region satisfy the requirement that $(1-\alpha)\ln\{q(t)\} \ge 0$. In these cases, positive gain is assured for all $K$ values since negative $K$ values are infeasible.\label{fig:gainK}}
	\end{figure}
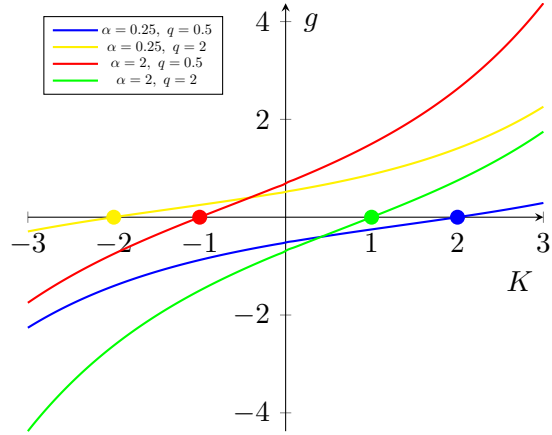

	The basic results of this section can be seen in Figure \ref{fig:gainK} for some specific parameter values.
	
	\subsection{GSLS under Brownian Motion Price Evolution\label{sec:GBM}}
	
	We now show that the GSLS strategy is robust in the case where the price evolution of the equity is determined by Geometric Brownian Motion (GBM) such that
	\begin{equation*}
	\frac{dp}{p} = \mu \ dt + \sigma \ dW
	\end{equation*}
	where $W$ represents a Weiner process, $\mu$ is the drift and $\sigma$ is the volatility \cite{hull2016options, oksendal2003stochastic}. Our proof follows along similar lines to that of \cite{barmish2016new}.
	
	As in \cite{barmish2016new}, the gain made by the long controller is
	\begin{equation*}
	g_L(t) = \frac{I_0}{K}\left\{q(t)^{K}e^{\frac{1}{2}\sigma^2(K-K^2)t} - 1\right\}
	\end{equation*}
	and, similarly, the gain made by the short controller is
	\begin{equation*}
	g_S(t) = \frac{\alpha I_0}{\beta K}\left\{q(t)^{-\beta K}e^{\frac{1}{2}\sigma^2(-\beta K-(\beta K)^2)t} - 1\right\}
	\end{equation*}
	and, hence, the total gain is
	\setlength{\arraycolsep}{0.0em}
	\begin{eqnarray*}
		g(t)&{}={}&\frac{I_0}{K}\left[ q(t)^{K}e^{\frac{1}{2}\sigma^2(K-K^2)t} - 1 \right.\nonumber\\
		&&\left. + \frac{\alpha}{\beta}\left\{q(t)^{-\beta K}e^{\frac{1}{2}\sigma^2(-\beta K-(\beta K)^2)t} - 1\right\}\right].
	\end{eqnarray*}
	\setlength{\arraycolsep}{5pt}
	The expected gain can then be derived by noting that the $k$th moment of a log-normally distributed random variable $X$ with $\log(X) \sim \mathcal{N}\Big(\big(\mu - \frac{\sigma ^2}{2}\big)t, \ \ \sigma^2 t\Big)$ is given by
	\begin{equation*}
	\mathbb{E}\left[X^k\right] \ = \ e^{\big(k\mu - \frac{\sigma ^2}{2}\big)t + \frac{1}{2}k^2 \sigma^2 t}.
	\end{equation*}
	Using this result, it can be shown that
	\begin{equation}
	\mathbb{E}\left\{g(t)\right\} \ = \frac{I_0}{K}\left\{e^{K\mu t} - 1 +\frac{\alpha}{\beta}\left(e^{-\beta K \mu t} - 1\right) \right\}.
	\label{Eg}
	\end{equation}
	The expected gain function under GBM, (\ref{Eg}), is of the same form as that for deterministic price evolution, (\ref{ComboEq}), with $q(t) = e^{\mu t}$. Thus, Theorem 1 follows immediately and, hence, positive expected gain follows when $(1 - \alpha) \mu \ge 0$. Similarly, Theorem 2 applies so that the (\eqref{Eg}) increases in $K$ and $\beta$.
	
	Note that the variance of the gains is given by
	\red{
		\begin{align}
		\mathrm{Var}\left\{g(t)\right\} &=
		\mathbb{E}\left\{g(t)^2\right\} - \left[\mathbb{E}\left\{g(t)\right\}\right]^2 \notag\\
		&= \frac{I_0^2}{K^2}\Bigg\{e^{\ 2K\mu t}\left(e^{\ \sigma^2 K^2 t} - 1\right) +\frac{\alpha}{\beta}e^{\ -\beta K \mu t}\Bigg[\frac{\alpha}{\beta}e^{\ -\beta K \mu t} \notag \\
		&\qquad\Big(e^{\beta^2K^2\sigma^2t}- 1\Big) + 2e^{K\mu t}\left(e^{-\beta K^2 \sigma^2 t} - 1\right)\Bigg]\Bigg\}
		\label{VarG}
		\end{align}
		Clearly the form of the variance is quite complicated even for this simple GBM model. However, some insights can be gained from Figure \ref{fig:gslsvar} where we consider the variance as a function of $\sigma^2$ and $\mu$ for various control parameter settings (and note that this generalizes Figures 4 and 5 of \cite{baumann2017stock}). We find that the variability increases in $\sigma^2$, and asymmetrically in $|\mu|$ (but is symmetric for the SLS case, i.e., when $\alpha=\beta=1$). In terms of the control parameters, it appears that the variance increases in $K$ and $\beta$, but depends on $\alpha$ in a more complex way. The key message here is that both $\mathbb{E}{\{g(t)\}}$ and $\mathrm{Var}{\{g(t)\}}$ depend on the control parameters, and, for example, simply increasing $K$ is not satisfactory in practice as it can yield highly variable returns (as noted by \cite{barmish2016new}  and \cite{baumann2017stock}). Thus, we suggest that a practical optimization procedure for selecting control parameters should take account of both the mean and the variance of gains.
	}
	
	\begin{figure}[h]
		\begin{subfigure}[b]{0.49\textwidth}
			\centering
			\begin{tikzpicture}[scale = 0.95]
			\pgfmathdeclarefunction{var}{6}{\pgfmathparse{%
					(1/#1^2)*(exp(2*#1*#2*#3)*(exp(#4*(#1^2)*#3)-1) + (#5/#6)*exp(-#6*#1*#2*#3)*((#5/#6)*exp(-#6*#1*#2*#3)*(exp((#6^2)*(#1^2)*#4*#3) -1) + 2*exp(#1*#2*#3)*(exp(-#6*(#1^2)*#4*#3) - 1)))}}
			\begin{axis}[unbounded coords=discard, domain = 0:0.2, axis lines = center, xlabel=$\sigma^2$, ylabel = $\mathrm{Var}\left\{g(t)\right\}$,
			x label style={at={(axis description cs:1.1,0.01)},anchor=east},
			y label style={at={(axis description cs:0.25,0.95)},anchor=east},
			tick align=outside,legend style={at={(.05,0.77), nodes={scale=0.5, transform shape}}, anchor=north west,legend columns = 1, font = \tiny}, every y tick scale/.style={at={(yticklabel cs:0.5)}, anchor = south, rotate = 90}, y tick scale label style = {at ={(0.52,0.9)}}]
			\addplot[samples = 100, blue, smooth, thick] {var(1,0.1,1,x,0.5,1)};
			\addlegendentry{$\alpha = 0.5, \beta = 1$, $K = 1$}
			\addplot[samples = 100, blue, dashed, thick] {var(3,0.1,1,x,0.5,1)};
			\addlegendentry{$\alpha = 0.5, \beta = 1$, $K = 3$}
			\addplot[samples = 100, blue, dotted, thick] {var(1,0.1,1,x,0.5,3)};
			\addlegendentry{$\alpha = 0.5, \beta = 3$, $K = 1$}
			\addplot[samples = 100, red, smooth, thick] {var(1,0.1,1,x,1,1)};
			\addlegendentry{$\alpha = 1, \beta = 1$, $K = 1$}
			\addplot[samples = 100, red, dashed, thick] {var(3,0.1,1,x,1,1)};
			\addlegendentry{$\alpha = 1, \beta = 1$, $K = 3$}
			\addplot[samples = 100, red, dotted, thick] {var(1,0.1,1,x,1,3)};
			\addlegendentry{$\alpha = 1, \beta = 3$, $K = 1$}
			\end{axis}	
			\end{tikzpicture}
			\caption{$\mu = 0.1$}
		\end{subfigure}
		\begin{subfigure}[b]{0.49\textwidth}
			\centering
			\begin{tikzpicture}[scale = 0.95]
			\pgfmathdeclarefunction{var}{6}{\pgfmathparse{%
					(1/#1^2)*(exp(2*#1*#2*#3)*(exp(#4*(#1^2)*#3)-1) + (#5/#6)*exp(-#6*#1*#2*#3)*((#5/#6)*exp(-#6*#1*#2*#3)*(exp((#6^2)*(#1^2)*#4*#3) -1) + 2*exp(#1*#2*#3)*(exp(-#6*(#1^2)*#4*#3) - 1)))}}
			\begin{axis}[unbounded coords=discard, domain = -0.2:0.2, axis lines = center, xlabel=$\mu$, ylabel = $\mathrm{Var}\left\{g(t)\right\}$, 
			x label style={at={(axis description cs:1.1,-0.01)},anchor=east},
			y label style={at={(axis description cs:0.75,0.95)},anchor=east},
			tick align=outside,legend style={at={(-.05,1), nodes={scale=0.5, transform shape}}, anchor=north west,legend columns = 1, font = \tiny},
			ytick scale label code/.code={\pgfmathparse{int(-#1)}$\mathrm{Var}\left\{g(t)\right\} \cdot 10^{\pgfmathresult}$},
			every y tick scale/.style={at={(yticklabel cs:0.5)}, anchor = south, rotate = 90}, y tick scale label style = {at ={(0.52,0.9)}}]
			\addplot[samples = 100, blue, smooth, thick] {var(1,x,1,0.1,0.5,1)};
			\addplot[samples = 100, blue, dashed, thick] {var(3,x,1,0.1,0.5,1)};
			\addplot[samples = 100, blue, dotted, thick] {var(1,x,1,0.1,0.5,3)};
			\addplot[samples = 100, red, smooth, thick] {var(1,x,1,0.1,1,1)};
			\addplot[samples = 100, red, dashed, thick] {var(3,x,1,0.1,1,1)};
			\addplot[samples = 100, red, dotted, thick] {var(1,x,1,0.1,1,3)};
			\end{axis}	
			\end{tikzpicture}
			\caption{$\sigma^2 = 0.1$}
		\end{subfigure}
		\caption{\red{GBM model-based $\mathrm{Var}{\{g(t)\}}$ as a function of (a) $\sigma^2$ (with $\mu = 0.1$) and (b) $\mu$ (with $\sigma^2 = 0.1$) for GSLS with $\alpha = 0.5$, $\beta = K = 1$ (blue, solid), and $\alpha = \beta = K = 1$ (red, solid). Both of these basic strategies are varied by setting $K=3$ (dashed) and $\beta=3$ (dotted). Note that there are two SLS cases (i.e., when $\alpha=\beta=1$). \label{fig:gslsvar}}}
	\end{figure}
	
	\red{We note that, although the GBM model is considered here, this specific model does not play a major role in the context of this paper. Rather, it is an example model that can be used as a vehicle for producing $\mathbb{E}{\{g(t)\}}$ and $\mathrm{Var}{\{g(t)\}}$ --- and these quantities form the basis of our proposed parameter optimization in Section \ref{Optim}. Nonetheless, this GBM model is a useful starting point within our proposal (and is a popular model in practice), but, for example, in the SLS setting, \cite{primbs2017robustness} considered processes with time-varying dynamics whereas \cite{baumann2017stock} considered jump processes. Thus, the work of this article could be extended to such processes, or, indeed, one could make use of model-free estimates of mean and variance as discussed in Section \ref{sec:discussion}.}

	\section{Selection of Parameters\label{Optim}}
	
	While results such as those in the previous section provide insight into the behaviour of feedback-based strategies (especially the conditions leading to positive gain), they fall short of yielding a practical implementation in the sense of suggesting values of the control parameters in real applications. Therefore, in this section, we focus on possible criteria that one may optimize in order to select the control parameters within GSLS (and, hence, also SLS).
	
	As a first step we will make the assumption that the price evolution process can be modelled by GBM since this assumption yields explicit solutions for both $\mathbb{E}\left\{g(t)\right\}$ and $\mathrm{Var}\left\{g(t)\right\}$, and our suggested criteria will be based on these quantities; of course other price evolution models could be used. Now, let $g^*_t$ be some prespecified target gain for time-point $t$, and define the trading ``bias'' as
	\begin{equation}
	\text{bias}\left\{g(t)\right\} = \mathbb{E}\left\{g(t)\right\} - g^*_{t}\label{eq:bias}
	\end{equation}
	which is the expected difference between the realized gain, $g(t)$, and the target gain, $g^*_t$. Under the GBM assumption, this quantity will depend on the parameter $\mu$ as well as the control parameters, $\alpha$, $\beta$, and $K$. Therefore, as a first step, the GBM parameters can be estimated using standard inference procedures \cite{ait2002maximum}, and, following this, the control parameters can be selected by minimizing, $\left[\text{bias}\left\{g(t)\right\}\right]^2$, i.e., these are the control parameters which minimize the difference between $\mathbb{E}\left\{g(t)\right\} \text{and} \ g^*_{t}$.
	
	The suggestion above does not take account of the volatility of the stock price, as evidenced by the fact that the objective function does not depend on $\sigma$. Thus, an alternative criterion would involve both the so-called bias and the variability in gains since the latter was also identified as important in \ref{sec:GBM}. We therefore propose the trading ``mean squared error'' (MSE) as
	\begin{equation}
	\text{MSE}\left\{g(t)\right\} = \mathbb{E}\left\{g(t) - g^*_t\right\}^2 = \left[\text{bias}\left\{g(t)\right\}\right]^2 + \text{Var}\left\{g(t)\right\}\label{eq:mse}
	\end{equation}
	which depends on both $\sigma$ (via the variance term) in addition to $\mu$, and can be minimized with respect to the control parameters. Such control parameters might result in $\mathbb{E}\left\{g(t)\right\} < g^*_t$, but where the lower variation in gains justifies the choice, i.e., control parameters selected via $\text{MSE}\left\{g(t)\right\}$ will, in any given run, tend be closer to the target gain due to the lower variability in gains. \red{It is worth highlighting, that, although this proposed optimization procedure selects control parameters with variance in mind, recall that the controller itself is placed only on the gain function; the extension of additionally including a variance controller is also of interest but beyond the scope of this paper.}
	
	\section{Testing GSLS with Optimized Control Parameters\label{sec:testing}}

	\subsection{Simulation Study\label{sec:sim}}
	
	Before testing our proposed methods on real data, we first carry out a simulation study. We simulated stock prices according to GBM with drift $\mu = 0.1$, and a range of different volatility values, $\sigma = \{0.05, 0.1, 0.2\}$. Since $I_0$ simply plays the role of scaling the gain up and down, we fix $I_0 = 1$, and optimize both (\ref{eq:bias}) and (\ref{eq:mse}) with respect to the control parameters wherein the known values of the GBM parameters are inserted prior to this optimization. Furthermore, we assume that, after one year of trading (over 252 days), the trader is aiming for a 15\% return, i.e, $g^*_1 = 0.15$, so that he/she beats the market drift ($\mu=0.1$) by 5 percentage points.
	
	Although the two objective functions can be optimized using standard optimization algorithms, since there are only three parameters $(K, \ \alpha, \ \beta)$, we performed a discrete grid search over the parameter space. In particular, we used 10 equally spaced values for each parameter with $K \in [0, 5]$, $\alpha \in [0, 5]$, and $\beta \in [0, 5]$ yielding 1000 parameter combinations. For each of the 3 simulation scenarios, we found the optimal control parameters according to both bias and MSE minimization, and then applied the resulting strategy over 252 trading days for each of 1000 simulated GBM trajectories.
	
	\begin{figure*}
		\begin{subfigure}{0.5\textwidth}
			\centering
			\includegraphics[width=\linewidth]{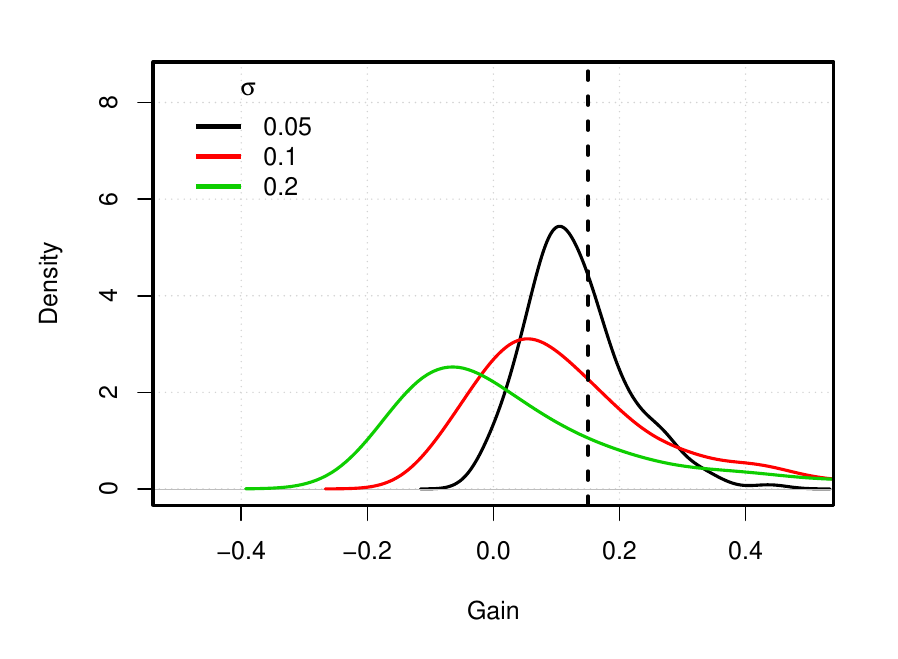}%
			\label{sfig:BiasDensity}
			\caption{Bias Optimization}
		\end{subfigure}
		\begin{subfigure}{0.5\textwidth}
			\centering
			\includegraphics[width=\linewidth]{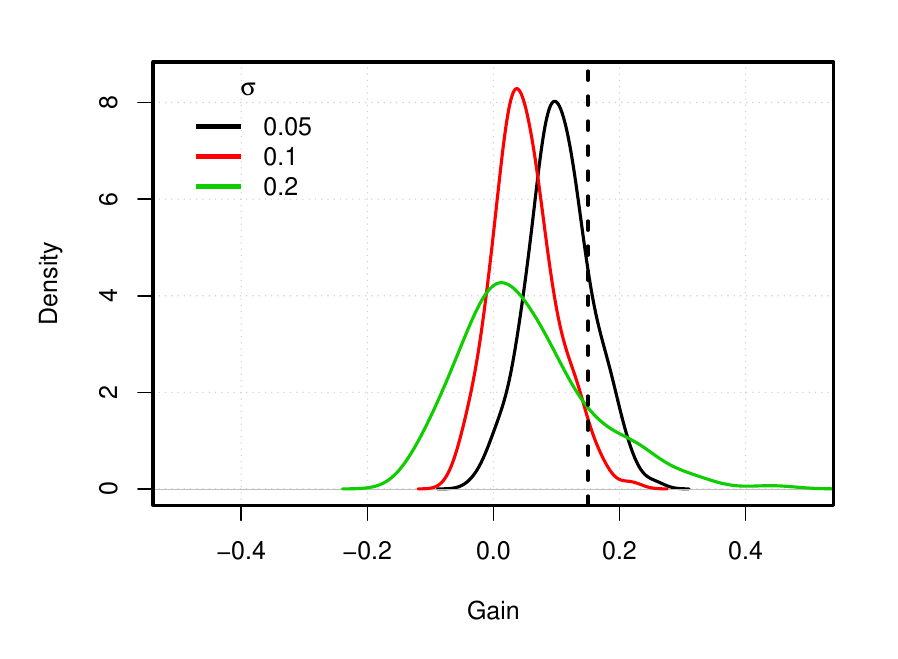}%
			\label{sfig:MSEDensity}
			\caption{MSE Optimization}
		\end{subfigure}
		\caption{Density of gains over 1000 simulated GBM stock prices over one year where control parameters were selected on the basis of (a) bias and (b) MSE. The target gain here is $g^*_t = 0.15$ and is shown by the dashed black line. \label{fig:biasmse}}
	\end{figure*}
	
	Figure \ref{fig:biasmse} shows the results based on optimizing both bias and MSE in terms of a density plot of the realized gains over each of the 1000 simulation replicates. We can see from Figure \ref{fig:biasmse}(a) that optimizing the bias does not appear to perform well as the volatility increases, \red{with much of the probability mass being on negative gains when $\sigma = 0.1$ and $\sigma = 0.2$; interestingly, however, the highly-right-skewed distributions also possess reasonable mass at large positive gain values.  By contrast, minimizing MSE (see Figure \ref{fig:biasmse} (b)) results in more consistent results as evidenced by the much tighter distribution of gains, and, although the distribution shifts towards negative gains for higher volatility (just like the bias optimization case), there is much less mass on negative gains. The tighter gains distributions do reduce the possibility of some potentially higher gains, but this is the classical mean-variance trade-off. These results suggest that parameter selection using MSE as an objective may be preferable for a trader who is also concerned about variability in gains in addition to the expected gains. Of course, a trader could weight the variance contribution in the MSE optimization to reflect their specific level of risk aversion, or, indeed, optimize bias subject to a fixed maximum variance in an ``efficient frontier'' manner \cite{markowitz1952portfolio}; we do not consider that here.}
	
	
	

	\subsection{S\&P 500 Stock Prices}
	
	As mentioned, in the existing literature, the degree to which feedback-based trading strategies have been tested on real data has been somewhat limited, i.e., the number of stock price series has been very small, and the choice of control parameters has been apparently quite arbitrary. Thus, more extensive testing is required if real-world traders are to be convinced that the adoption of such strategies can be fruitful in general.
	
\begin{figure}
		\centering
		\includegraphics[width = 0.5\textwidth]{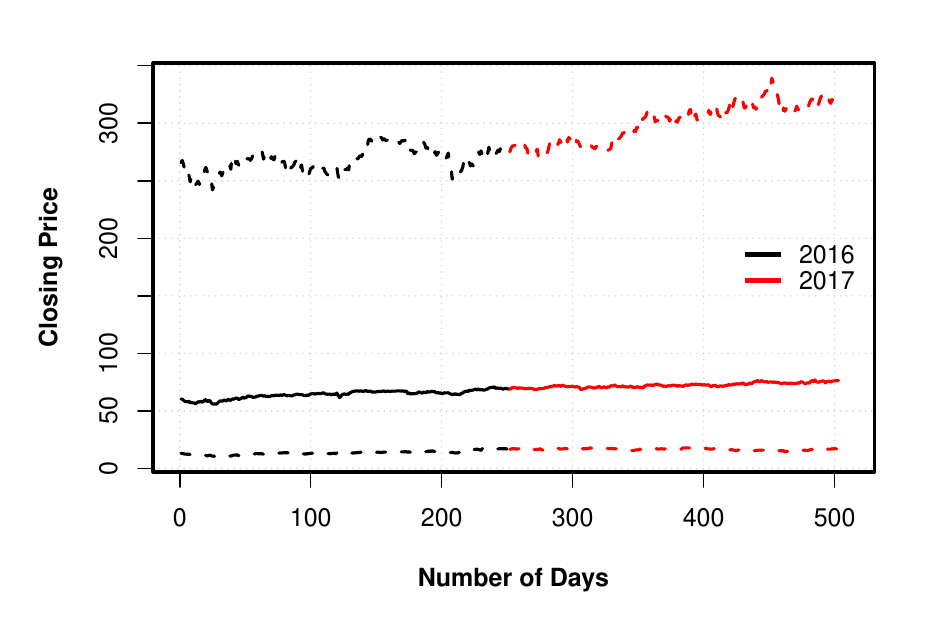}
		\caption{The 2.5\% (bottom, dashed), 50\% (middle, solid), and 97.5\% (top, dashed) quantiles of closing prices for 495 members of the S\&P 500 over the period January 2016 - December 2017.}
		\label{SP500}
	\end{figure}

	With the above goal in mind, we consider the daily closing prices for 495 members of the Standard and Poor's 500 (S\&P 500) index over the course of two-year period, January 2016 - December 2017 (of the current 500 members, it was not possible to obtain complete data for the time period in question in 5 cases which were, therefore, omitted).  
	The justification for choosing this data is based on the fact that the members of the S\&P 500 are the largest companies traded in the United States and eligibility is based on a number of factors including market capitalization\footnote{Market capitalization of \$5.3bn or more.}, liquidity\footnote{Annual dollar value traded to float-adjusted market capitalization should be greater than 1, and the stock should trade a minimum of 250,000 shares in each of the six months leading up to the evaluation date.}, and the company being domiciled\footnote{The company must be a U.S company as defined by the US Index Committee.} \cite{SP500}. Thus, these members have highly liquid stocks so that feedback-based trading strategies could plausibly be implemented in a real-world scenario, and, moreover, the members of this index are intended to provide a reasonably good representation of the stock market as a whole. 
	Figure \ref{SP500} shows the median adjusted closing price calculated over all 495 members on each day over the two-year period, along with 2.5\%, and 97.5\% quantiles; it is clear that the majority of the stock prices have increased over time.


	For each of the stock price series, we take the year 2016 to be the training period for which the GBM parameters are estimated using maximum likelihood \cite{ait2002maximum}, and then, based on the estimated GBM parameters, we select the GSLS control parameters via the bias and MSE optimization approaches described in Section \ref{Optim} with a target gain of $g^* = 0.15$ for each stock. Following Section \ref{sec:sim}, when optimizing GSLS control parameters, we use a grid search over 1000 parameter combinations resulting from 10 equally spaced values for each parameter, $K \in [0, 5]$, $\alpha \in [0, 5]$, and $\beta \in [0, 5]$. The optimized trading strategy (either bias and MSE) for that stock price series is then applied over the course of the year 2017, i.e., the testing period, where a discrete-time version of the strategy is implemented (see~\cite{barmish2016new}). While the procedure as described is carried out separately on each stock price series, we will summarize results by way of aggregating over all series. Note: in all of our testing, we fix $I_0=1$ for each stock since this simply scales the gains.
	
\begin{figure}[h!]
		\centering
		\includegraphics[width = 0.5\textwidth]{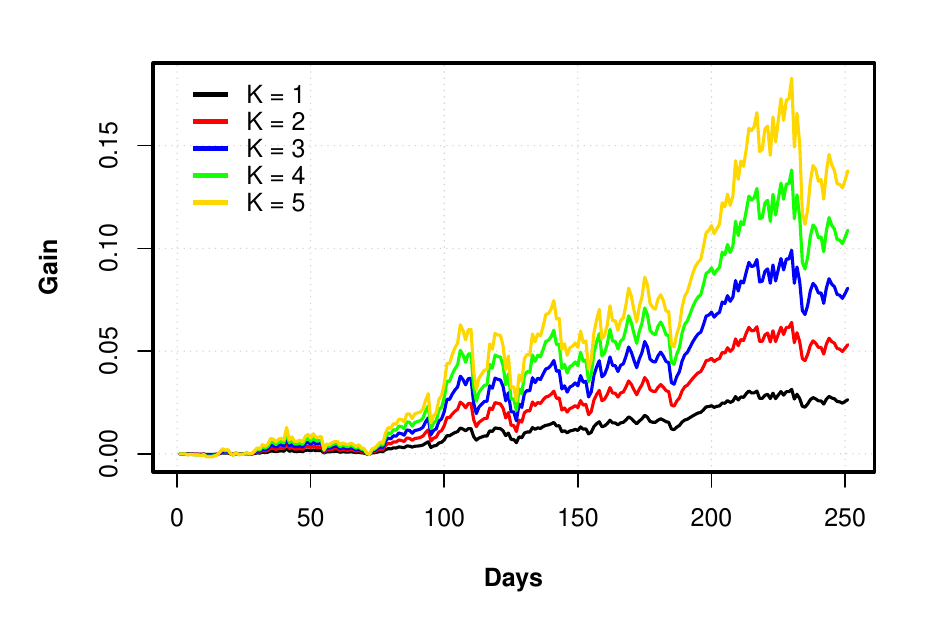}
		\caption{Average gain (computed over all stock series) on each trading day in 2017 from the SLS strategy with various $K$ values.}
		\label{SLSgains}
	\end{figure}

	Before applying any optimized strategies, we first test the classical SLS strategy in a similar way to previous literature, i.e., a value for the control parameter is simply selected and tested; we use $K = \{1,2,3,4,5\}$. The advantage here, however, is that we are testing over a much larger sample of price series than that of previous literature. So that the results are comparable with the case where we carry out optimization, we test these five strategies in the year 2017 only (since $K$ is simply fixed from the offset, 2016 plays no role in parameter selection). Figure \ref{SLSgains} displays the average gain on each trading day (average was computed over the 495 stock series). We can see that the average gain increases with the value of $K$, however the trajectory of the gain is quite erratic over time \red{(as expected from Section \ref{sec:GBM}  where we found that the variability increases with $K$)}; most of the growth appears as a sharp jump occurring in the last 50 days.
	
\begin{figure}[h!]
		\centering
		\includegraphics[width = 0.5\textwidth]{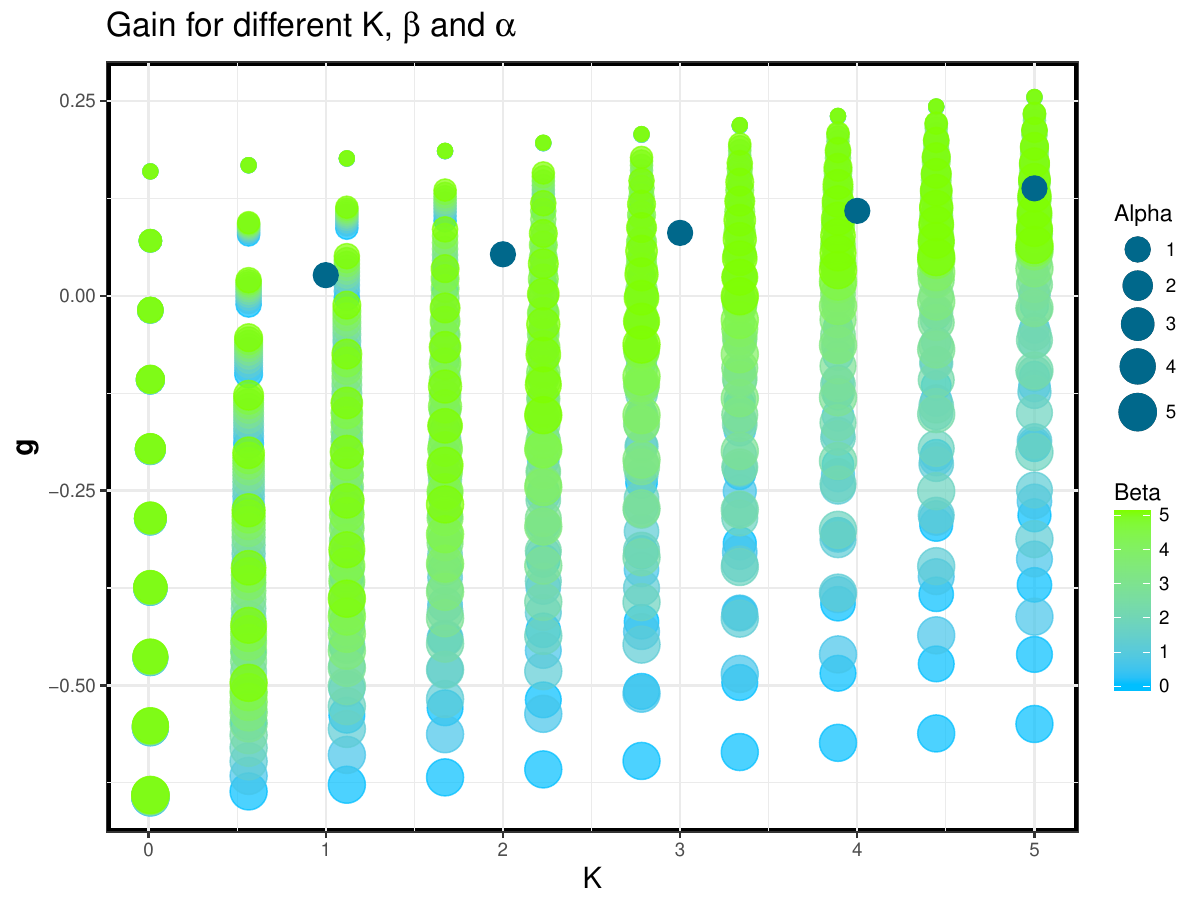}
		\caption{Average gain (computed over all stocks) after 252 days of trading in 2017 for each of 1000 GSLS strategies. The colour of the points representw the $\beta$ value, while their size represents the $\alpha$ value. Also shown are five SLS strategies (the dark coloured points).}
		\label{ABK}
	\end{figure}

	We then applied each of the 1000 GSLS parameter combinations to all stocks in the year 2017 (again without using 2016 to optimize), and computed the average gain over all stocks at the end of the year; the results are visualized in Figure \ref{ABK} along with the five values from the five SLS strategies mentioned in the previous paragraph. Interestingly, there are a large number of GSLS strategies which lead to a loss. These mainly correspond to cases where $\alpha > 1$ which is not a surprise since most of these stock prices are increasing on average over time as evidenced by Figure \ref{SP500}. Note also that, while the classic SLS case does outperform these particular combinations, there are clearly many other GSLS cases which outperform SLS.
	
\begin{figure}[h!]
		\centering
		\includegraphics[width = 0.5\textwidth]{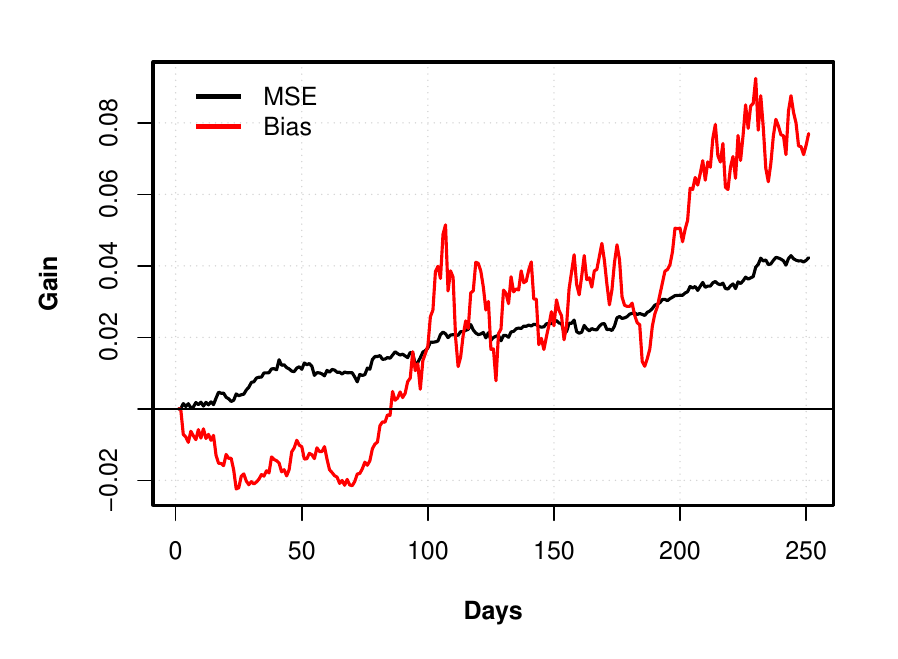}
		\caption{Average gain (computed over all stocks) on each trading day in \red{2017} for optimized strategies. Both the bias-optimized (red) and MSE-optimized (black) strategies were based on the data from 2016 and with $g^* = 0.15$.}
		\label{total_gains}
	\end{figure}

\begin{figure}[h!]
		\centering
		\includegraphics[width = 0.5\textwidth]{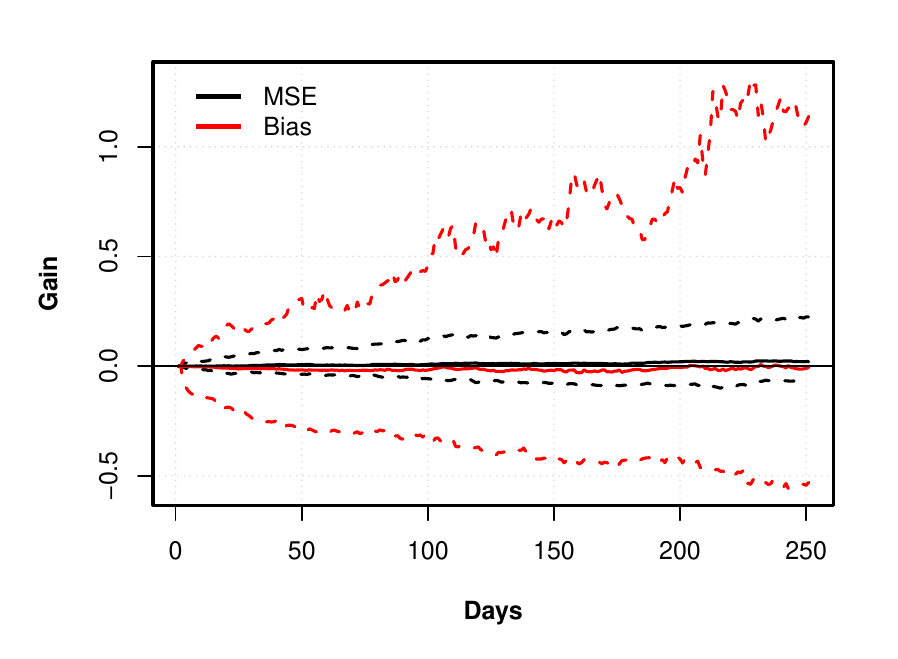}
		\caption{Median gain (computed over all stocks) for both the bias-optimized (red, solid) and MSE-optimized (black, solid) strategies, along with 2.5\% and 97.5\% quantiles (dashed) with $g^* = 0.15$.}
		\label{gain_quantiles}
	\end{figure}

	The tests described above (both SLS and GSLS) mimic those of previous literature, i.e., we have simply set the control parameters at the start of 2017 without drawing insight from historical data. In other words, the parameters could be thought of as essentially being randomly selected, and are not tuned on a per-stock basis. It is unlikely that a real-world trader would adopt such a strategy from which he/she is quite detached. We, therefore, implement our optimization approach for each stock based on the 2016 data (as described above), and apply the optimized strategies to the 2017 data. Figure \ref{total_gains} shows the average gain using optimized strategies (both bias and MSE) over each trading day, \red{where, in the end, we find that, while the bias approach performs better when measured purely in terms of gains, there is clearly much greater variability in comparison to the MSE approach which increases much more steadily over time (and note that the MSE case yields positive gains at all time points here).} To get a \red{further} sense of the variability in gains, Figure \ref{gain_quantiles} displays the 2.5\% and 97.5\% quantiles for the gains over all the stocks. Unsurprisingly, the level of variability associated with the bias-optimized strategies is \red{significantly} higher than that of the MSE-optimized strategies. \red{Clearly there is a much greater chance of negative gains in the bias-optimized strategy, but the right-skewed distribution does achieve high gains some ``lucky'' cases. On the other hand, the MSE-optimized strategy is more consistent in the sense that the gains distribution is much tighter, and with less chance of negative gains; this } is in line with the findings of Section \ref{sec:sim}. 

	%
	%
	%
	
	\begin{figure}[h!]
		\centering
		\includegraphics[width = 0.5\textwidth]{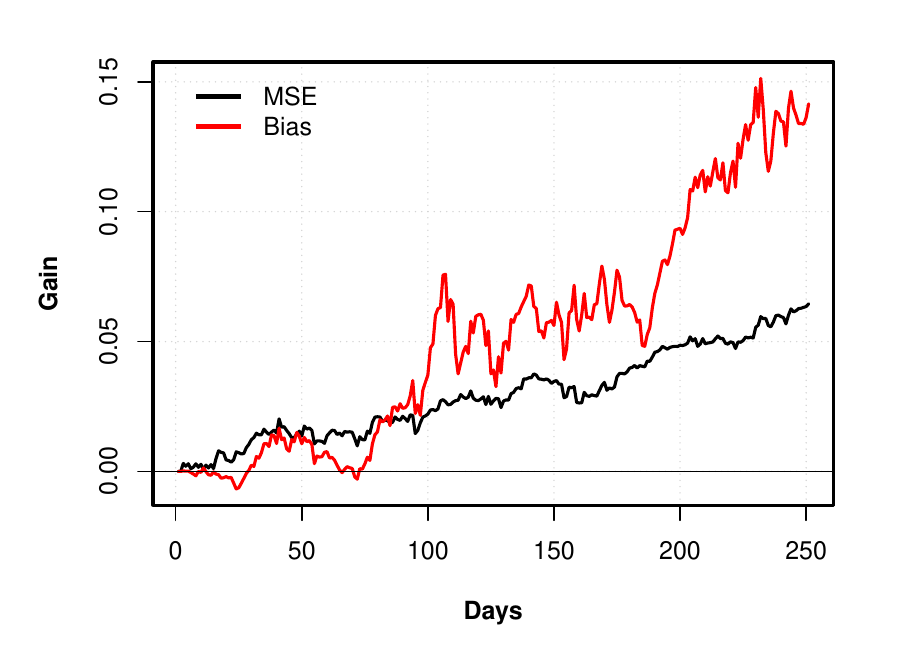}
		\caption{Average gain (computed over all stocks) on each trading day in \red{2017} for optimized strategies with $g^* = |\mu| + 0.05$. Both the bias-optimized (red) and MSE-optimized (black) strategies were based on the data from 2016.}
		\label{mean_DiffTargetG}
	\end{figure}
	
	In the optimization procedures discussed above, the target gain was fixed at 15\% for all stocks. However, the GBM parameters could also be used in order to determine a reasonable target gain, for example, a target of 15\% might be an unobtainable level of gain for some of the stocks in question, or, indeed, an underestimate of the potential gain for other stocks. Thus, one might consider varying the target on a stock-by-stock basis. A simple suggestion in this direction would be to set the target gain, for stocks $i=1,\ldots,495$, at $g_{t,i}^* = |\hat\mu_i| + C$ where $\hat\mu_i$ is the estimated GBM drift parameter and $C$ is some constant which describes the amount by which we wish to beat the inherent drift in the price evolution. The reason we take the absolute value of the drift is that profit can also be made in the case of negative drift via the short trading component. Figure \ref{mean_DiffTargetG} displays the average gain over time with $C = 0.05$. While the average gain in the MSE-optimized case \red{increases by approximately 50\% compared to that seen in Figure \ref{total_gains} (final gain goes from $0.042$ to $0.065$)}, the average gain for the bias-optimized case is almost doubled \red{(from $0.077$ to $0.142$)}. This finding is perhaps to be expected since the bias optimization is solely chasing a target, whereas the MSE approach also takes account of variability, \red{i.e., altering the target has a bigger impact on the bias-optimizer.}

	\begin{table*}[htb]
		\centering
		\caption{Summary of gains over all stock at the end of the year 2017 using different strategies}
		\label{Gtable}
		\renewcommand{\arraystretch}{1.3}
		\resizebox{\textwidth}{!}{\begin{tabular}{|cr|rr|rr|rrr|rr|rr|c|}
				\hline
				& Strategy		& \multicolumn{4}{|c|}{\textbf{GSLS}} & \multicolumn{7}{|c|}{\textbf{SLS}} \\
				\hline
				& Parameters   & MSE    & MSE     & Bias   & Bias   & $K=1$    & $K=2$    & $K=5$    & MSE      & MSE      & Bias    & Bias \\
				\hline
				& Target       & Fixed  & Varied  & Fixed  & Varied & ---      & ---      & ---      & Fixed    & Varied   & Fixed   & Varied  \\
				\hline
				\multirow{3}{*}{Summary} &		1st Quartile    &  \red{-0.018} &  \red{-0.018} & -0.158 & -0.154 & -0.029   & -0.057   & -0.131   & \red{$|<\!0.001|$} & \red{-0.002}   & -0.083 & -0.114 \\
				\multirow{3}{*}{of}      &            Median    &  \red{0.021} & \red{0.037} & -0.006 &  0.026 &$<\!0.001$& -0.002   & -0.020   &\red{$<\!0.001$} & \red{$<\!0.001$} & -0.006 & -0.017 \\
				\multirow{3}{*}{Gains}   &           Mean    &  \red{0.042} &  \red{0.065}  &  0.077 &  0.142 &  0.026   &  0.053   &  0.138   &  \red{0.007}   &  \red{0.013}   &  0.108 &  0.124 \\
				&      3rd Quartile    &  \red{0.098} &  \red{0.126}  &  0.185 &  0.264 &  0.060   &  0.115   &  0.237   &  \red{0.007} &  \red{0.011}   &  0.128 &  0.217 \\
				& Inter-Quartile Range & \red{0.116}  &  \red{0.144}  &  0.343 &  0.418 &  0.089   &  0.172   &  0.368   &  \red{0.002}   &  \red{0.013} &  0.211 & 0.331 \\
				& \red{Standard Deviation} & \red{0.082} & \red{0.142} & \red{0.448} & \red{0.455} & \red{0.102} & \red{0.204} & \red{0.520} & \red{0.026} & \red{0.047} & \red{0.425} & \red{0.461} \\
				& \red{Loss Probability} & \red{0.402} & \red{0.383} & \red{0.513} & \red{0.430} & \red{0.499} & \red{0.503} & \red{0.533} & \red{0.497} & \red{0.497} & \red{0.520} & \red{0.525} \\
				
				\hline
		\end{tabular}}
		\vspace{0.2cm}
		
		\footnotesize{``MSE'' and ``Bias'' indicate that parameters were selected by optimizing these criteria for each stock based on 2016 data where ``Fixed'' means that $g^* = 0.15$ and ``Varied'' means $g^* = |\mu| + 0.05$. Inter-Quartile Range $=$ 3rd Quartile $-$ 1st Quartile.}
	\end{table*}

	\red{Table \ref{Gtable} summarizes the gains distribution (over all stocks) for the following strategies: optimized GSLS (based on the MSE and bias objectives with both a fixed and varied target), SLS for some fixed $K$ values (which is essentially what appears in the literature so far), and optimized SLS (where optimization is done with respect to $K$ only since $\alpha = \beta = 1$). Although the MSE-optimized GSLS strategy does not have the largest mean gain of all strategies (as expected since it hedges against risk), it does appear to yield a good mean-variance compromise. The standard deviation of gains is reasonably low compared to the other strategies. This reduced variability also reduces the chance of large losses (which is high for other strategies), and the median gain is positive (this is negative or close to zero for most other strategies); interestingly, varying the target per stock increases the variability a little, but improves the level of gains. On the other hand, bias-optimized GSLS produces a larger mean gain but with quite high variability such that, in 25\% of cases, the loss is below $-0.158$. This strategy performs quite similarly to SLS with $K=5$ and also to bias-optimized SLS; of course, these latter two strategies are essentially the same since the gain increases in $K$. However, MSE-optimized SLS yields gains which are quite tightly packed around zero.
		
		In summary, it seems that MSE optimization provides a good criterion for selecting control parameters, but only in the case of GSLS. Optimizing the classical SLS does not yield a useful strategy as the method does not have sufficient flexibility to simultaneously increase the gain but reduce its variation with only a single parameter, $K$ (and classical SLS without parameter selection is not a satisfactory procedure). The Appendix contains analogous results for a different time frame wherein no strategy does well on average; however, MSE-optimized GSLS does reduce exposure to very large losses.}

	\section{Discussion\label{sec:discussion}}
	
	The GSLS strategy expands classical SLS so that the parameters of the long and short controllers can differ (i.e., the initial investment and feedback parameters). By allowing different initial long and short investments such that the overall net initial investment is nonzero, $I(0) \ne 0$, the GSLS strategy can be thought of as a paradigm lying between SLS and more standard trading strategies (e.g., simply going long or short). This provides the opportunity to include some insight into the trading strategy while lowering risk relative to the standard trading strategies.
	
	The robustness of GSLS with differing long and short initial investments ($\alpha\ne1$) is weakened relative to SLS in the sense that the theoretical (expected) gain is no longer guaranteed to be positive; interestingly, GSLS with differing feedback parameters ($\beta\ne1$) maintains robustness provided that $\alpha=1$. However, should the trader possess some knowledge of the price evolution, and select $\alpha\ne1$, a greater level of gain can be achieved than that of SLS. Indeed, the primary driver of any realized gain is, as one would expect, the price evolution of the stock itself -- which is why one might wish to make estimates about its likely evolution at the expense of some robustness.
	
	
	Throughout the current paper trader ``knowledge'' entered in the form of a GBM assumption of price evolution which is common in the financial literature \cite{black1973pricing, merton1974pricing}. Of course, this is a simplifying assumption, and, in practice, more general models could be used which could include other forms of market knowledge exogenous to the historical price series itself. The secondary reason for using the GBM model is that it admitted a closed form solution for the expected gain and the variance of the gain. In more complex models, these quantities could be obtained by simulation. The model assumption could potentially be avoided altogether by using some model-free estimate of $\mathbb{E}\left\{g(t)\right\}$ such as the last realized gain value achieved in a historical sequence $\{g_1, g_2, \ldots, g_n\}$. Another alternative might be a weighted average over these realized gains $\sum_{j=1}^n w_j g_j$ where the weights, $w_j$, grow with $j$ so that more weight is placed on more recent observations. In any case, in our practical application, the methods appeared to perform well despite the GBM assumption.
	
	Selection of control parameters which are ``optimal'' in some sense has not previously been considered in the literature to our knowledge. The lack of such selection procedures presents a major hurdle in the wider adoption of feedback-based trading strategies. To this end, we have proposed two possibilities which we call bias and MSE optimization. The bias optimization focusses on the expected gain, whereas the MSE approach additionally takes account of the variation of gains. Under the GBM assumption, the expected gain is connected to the drift of the process, while the variation incorporates the process volatility. While our suggestion involves specifying a target gain (for which a reasonable value can be informed by the estimated GBM model), we might, alternatively, have simply selected the best performing strategy over the testing period. This is equivalent to minimizing bias with some arbitrarily large target gain. Note also that our MSE objective function, whilst being a very natural quantity in itself, places equal weight on the bias and variance of the gain; more generally, a tuning parameter might control the trade-off between these two quantities.
	
	Another parameter selection approach still (akin to the so-called ``efficient frontier'' in portfolio optimization \cite{markowitz1952portfolio}) would be based on maximizing the expected gain subject to a fixed level of variance, or minimizing the variance subject to a fixed expected gain. We could also imagine an optimization procedure which updates throughout time such that the control parameters are altered dynamically (albeit the initial investments are obviously fixed from the start); as pointed out by an anonymous reviewer, this could make use of work due to \cite{li2000optimal} and \cite{zhou2000continuous} which considers a multi-period efficient frontier to analytically determine optimal portfolio weights by incorporating the trader's wealth through a linear feedback controller. Nevertheless, parameter selection is novel in the SLS context as previously discussed. Therefore, the procedure set out in this article provides a useful starting point, and it has performed well in our extensive testing.

	In the SLS setting, the effect of $K$ on the expected return, $\mathbb{E}\{g(t)\}$, as well as the risk, $\mathrm{Var}\{g(t)\}$, was noted previously by \cite{barmish2016new}  and \cite{baumann2017stock}. In our GSLS setting, we have highlighted the importance of incorporating both quantities into the control parameter optimization. However, we note that the controller itself is placed only on the return. Therefore, while this work goes beyond existing SLS literature by optimizing parameters with risk mitigation in mind (as discussed in the previous paragraph), it would be interesting to consider additionally placing a controller on the risk in a similar manner to \cite{li2000optimal} but incorporating both a long and short component; of course, that is beyond the scope of this paper.
	
	
	
	Overall, despite being a relatively new area of research, feedback-based trading strategies clearly have interesting theoretical properties as well as promising performance in practice. Through the extension of classical SLS to GSLS, the investigation of parameter selection procedures, and the extensive analysis of performance in a large sample of real stock price data, this paper contributes to a greater understanding of the potential of these strategies. \red{There are various avenues for potential research in this emerging area (some indicated in the above discussion), and we suggest that future developments could follow a similar standard of testing to that used in this paper (i.e., on many real stock price series).}

	\section*{Acknowledgment}
	The authors would like to thank James Gleeson for his helpful comments on an earlier draft of this paper.
	\bibliographystyle{IEEEtran}
	\bibliography{Bibliography}
	\newpage
	\vfill
	\appendix
	
	\section{Further Testing}\label{sec:Appendix}
\begin{figure}[h]
		\centering
		\includegraphics[width = 0.5\textwidth]{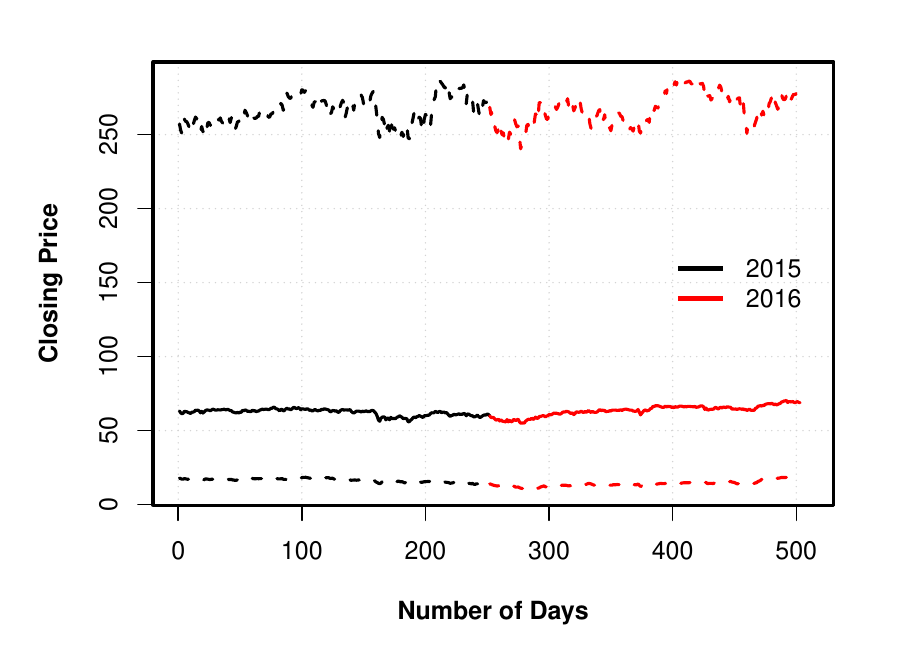}
		\caption{\red{The 2.5\% (bottom, dashed), 50\% (middle, solid), and 97.5\% (top, dashed) quantiles of closing prices for 461 members of the S\&P 500 over the period January 2015 - December 2016.}}
		\label{SP500_1516}
	\end{figure}

	Here we consider equivalent testing to that of Section \ref{sec:testing}, but over the period 2015-2016 during which the stock prices did not have a general upward trend as can be seen in Figure \ref{SP500_1516}. Note that the number of stocks used in this testing is 461 (rather than 495) due to the availability of data over this period.

	\begin{table*}[t!]
		\centering
		\color{black}
		\caption{Summary of gains over all stock at the end of the year 2016 using different strategies}
		\label{Gtable16}
		\renewcommand{\arraystretch}{1.3}
		\resizebox{\textwidth}{!}{\begin{tabular}{|cr|rr|rr|rrr|rr|rr|c|}
				\hline
				& Strategy		& \multicolumn{4}{|c|}{\textbf{GSLS}} & \multicolumn{7}{|c|}{\textbf{SLS}} \\
				\hline
				& Parameters   & MSE    & MSE     & Bias   & Bias   & $K=1$    & $K=2$    & $K=5$    & MSE      & MSE      & Bias    & Bias \\
				\hline
				& Target       & Fixed  & Varied  & Fixed  & Varied & ---      & ---      & ---      & Fixed    & Varied   & Fixed   & Varied \\
				\hline
				\multirow{3}{*}{Summary} &	1st Quartile & -0.031 & -0.047  & -0.200 & -0.222 & -0.056 & -0.109 & -0.223 & -0.003 & -0.024 & -0.165 & -0.194 \\
				\multirow{3}{*}{of} & Median & -0.016 & -0.018  & -0.087 & -0.097 &-0.026& -0.054 & -0.131 & $|<\!0.001|$ &$|<\!0.001|$ & -0.076 & -0.113 \\
				\multirow{3}{*}{Gains}   & Mean & 0.002 & -0.017  & -0.074 & -0.062 & -0.012 & -0.025 & -0.049 & -0.004 & -0.010 & -0.054 & -0.078\\
				& 3rd Quartile & 0.026 & 0.029 & 0.051 & 0.054 & 0.010 & 0.015 & -0.019 & $<\!0.001$ & $<\!0.001$ & -0.004 & -0.018\\
				& Inter-Quartile Range & 0.057 & 0.076 & 0.250 & 0.276 & 0.066 & 0.124 & 0.203 & 0.003 & 0.025 & 0.161 & 0.176\\
				& Standard Deviation & 0.057 & 0.134 & 0.287 & 0.357 & 0.154 & 0.298 & 0.910 & 0.015 & 0.053 & 0.257  & 0.237\\
				& Loss Probability & 0.636 & 0.633 & 0.692 & 0.677 & 0.688 & 0.692 & 0.792 & 0.685 & 0.685 & 0.770 & 0.777 \\
				
				\hline
		\end{tabular}}
		\vspace{0.2cm}
		
		\footnotesize{``MSE'' and ``Bias'' indicate that parameters were selected by optimizing these criteria for each stock based on 2015 data where ``Fixed'' means that $g^* = 0.15$ and ``Varied'' means $g^* = |\mu| + 0.05$. Inter-Quartile Range $=$ 3rd Quartile $-$ 1st Quartile.}
		\color{black}
	\end{table*}

\begin{figure}[t!]
		\begin{subfigure}[b]{0.5\textwidth}
			\centering
			\includegraphics[width=\linewidth]{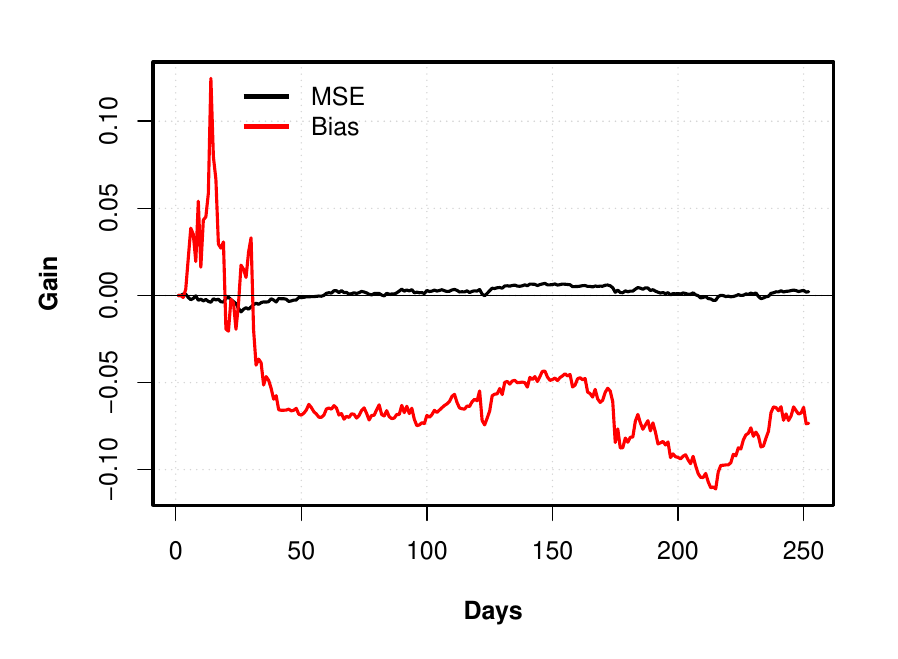}
			\caption{Fixed Target}
		\end{subfigure}
		\begin{subfigure}[b]{0.5\textwidth}
			\centering
			\includegraphics[width=\linewidth]{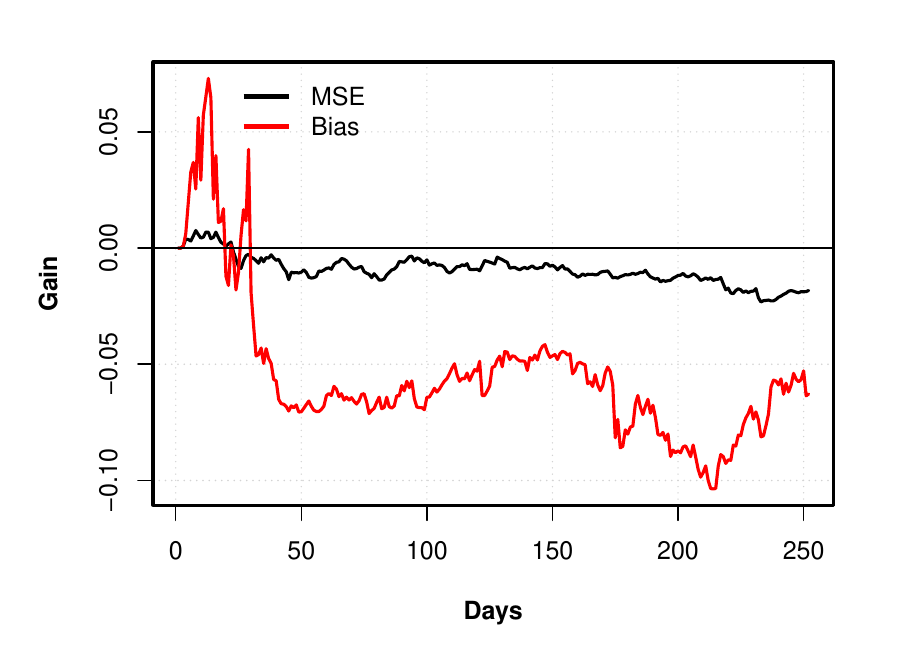}%
			\caption{Varied Target}
		\end{subfigure}
		\caption{Average gain (computed over all stocks) on each trading day in 2016 for optimized strategies with (a) $g^*=0.15$ and (b) $g^* = |\mu| + 0.05$. Both the bias-optimized (red) and MSE-optimized (black) strategies were based on the data from 2015.} \label{fig:biasmse1516}
	\end{figure}

	We now estimate parameters based on 2015, and test on 2016. Figure \ref{fig:biasmse1516} displays the average gain for optimized GSLS with fixed ($g^*_t = 0.15$) and varied ($g^*_t = |\mu| + 0.05$) targets, and summaries for various strategies can be found in Table \ref{Gtable16}. Although no strategy does well in this case, the MSE-optimized GSLS strategy at least reduces the risk of larger losses.
	\vfill

\end{document}